\documentstyle[12pt,amssymb,epsfig,a4]{article} 
\newcommand{\vs}{\vspace{-0.25cm}}
\begin{document} 

\begin{center}
{\Large{\bf Mean eigenvalues for simple, simply connected, compact Lie groups}}

\medskip

N. Kaiser\\

\smallskip

{\small Physik Department T39, Technische Universit\"{a}t M\"{u}nchen,
    D-85747 Garching, Germany

\smallskip

{\it email: nkaiser@ph.tum.de}}
\end{center}

\medskip

\begin{abstract}
We determine for each of the simple, simply connected, compact and complex Lie 
groups $SU(n)$, Spin$(4n+2)$ and $E_6$ that particular region inside the unit 
disk in the complex plane which is filled by their mean eigenvalues. We give 
analytical parameterizations for the boundary curves of these so-called trace 
figures. The area enclosed by a trace figure turns out to be a rational
multiple of $\pi$ in each case. We calculate also the length of the boundary 
curve and determine the radius of the largest circle that is contained in a 
trace figure. The discrete center of the corresponding compact complex Lie 
group shows up prominently in the form of cusp points of the trace figure
placed symmetrically on the unit circle. For the exceptional Lie groups $G_2$, 
$F_4$ and $E_8$ with trivial center we determine the (negative) lower bound on 
their mean eigenvalues lying within the real interval $[-1,1]$. We find the 
rational boundary values $-2/7$, $-3/13$ and $-1/31$ for $G_2$, $F_4$ and 
$E_8$, respectively.         
\end{abstract}

\bigskip
PACS: 02.20.Qs, 02.20.Rt. 
\section{Introduction and summary}
The dynamical understanding of the confinement-deconfinement phase transition 
as a function of the temperature in a nonabelian gauge theory, such as quantum 
chromodynamics, is a topic of current interest. The so-called Polyakov loop,
given by the trace of the thermal Wilson line, has been proposed as an order
parameter for the confinement-deconfinement transition \cite{svet}. Numerical
simulations of pure $SU(3)$ Yang-Mills gauge theory on an euclidian spacetime 
lattice have verified this proposal by computing the renormalized Polyakov 
loop as a function of the temperature \cite{latt} (see Fig.\,4 therein). 
Because of its special role as an order parameter of the 
confinement-deconfinement transition effective Lagrangian in the Polyakov loop 
variable have been formulated \cite{pisa}. There is also ongoing 
activity to interpret within such a framework the results of lattice QCD
simulations at finite temperature including dynamical quarks
\cite{dumitru,ratti}.  

From the mathematical point of view the Polyakov loop variable is a 
complex number given by 1/3  the trace of a special unitary $3\times 3$ 
matrix (i.e. ${1\over 3}{\rm tr}\,U,\, U\in SU(3)$). According to this
definition as an arithmetic mean of three unitary eigenvalues (with product 
equal to $1$) only restricted complex values are possible for the Polyakov
loop variable, and this feature should be respected in the construction of any
effective Lagrangian. We will show here in section 2 that the allowed region of
${1\over 3}{\rm tr}\,U,\, U\in SU(3)$ lies inside the unit circle and is
bounded by a quartic curve with three cusp points at the cube roots of unity:
$1,\, (-1\pm i \sqrt{3})/2$. Further geometrical properties of this so-called
trace figure of $SU(3)$, such as its enclosed area and the length of its
circumference will also be derived.  

The question about the domain of their mean eigenvalues naturally generalizes
to the other compact Lie groups with a complex fundamental representation.  The
simple and simply connected, compact Lie groups (without any abelian
$U(1)$-factors) are particularly interesting, since for these the trace figure
(i.e. locus of the mean eigenvalues) will not be the entire unit disk in the 
complex plane. In section 3 we analyze the special unitary groups $SU(n),\, n 
\geq 4$ and show that their trace figures are bounded by hypocycloids with
cusp points at the $n$-th roots of unity. Section 4 deals then with the spin 
groups Spin$(4n+2)$ (the two-sheeted covering groups of the special orthogonal 
groups $SO(4n+2)$) which possess a complex fundamental representation of
dimension $4^n$. The exceptional Lie group $E_6$ with its  complex fundamental 
representation of dimension 27 will be analyzed in section 5. For the compact 
Lie groups $G$ with only selfconjugate representations the mean eigenvalues
are necessarily real and restricted to the interval $[-1,1]$. If furthermore
the center of the group $Z(G)$ contains a factor ${\mathbb{Z}}_2$, it follows
by continuity that the whole interval $[-1,1]$ gets filled by the mean 
eigenvalues of $G$. This leaves for the exceptional Lie groups $G_2$, $F_4$
and $E_8$ with trivial center the question about a lower bound on their mean
eigenvalues. We will determine in section 6 these lower bounds for $G_2$,
$F_4$ and $E_8$ as the negative rational numbers $-2/7$, $-3/13$ and $-1/31$,
respectively.        
                
\section{Special unitary group SU(3)}
The simplest among the compact and complex Lie groups is the special unitary
group $SU(3)$, with numerous applications in theoretical physics. The question 
about the domain in the complex plane filled by the mean eigenvalues of
$SU(3)$-matrices is readily answered. The complex numbers of the form: 
\begin{equation}  X+i\,Y = {1\over 3}\, {\rm tr}\,U \,, \quad U \in SU(3) \,,
\end{equation} 
can be rewritten in terms of the three unitary eigenvalues $z_j = e^{i 
\theta_j}\in U(1)$ of $U$, which are subject to the constraint $z_1z_2z_3=1$, 
as: 
\begin{equation}  X+i\,Y = {1\over 3}\bigg(z_1+z_2 +{1\over z_1 z_2} \bigg)
  \,.\end{equation}
Evidently, any such complex number $X+i\,Y$ lies inside the unit disc: $X^2+
Y^2 \leq 1$.  In order to determine the extremal boundary values we eliminate
one of the two angles  by imposing the zero-derivative condition: $\partial(X+
i\, Y)/ \partial z_1 =0$, which gives $z_2= z_1^{-2}$. Reinserting this
relation leads already to the following parameterization of the boundary 
curve:   
\begin{equation}  3(X_b+i\,Y_b) = 2z_1 +z_1^{-2} = 2\cos^2 \theta +2\cos
\theta-1 +2i \, (1-\cos\theta) \sin\theta   \,,\end{equation}
where $\theta$ is an angle running from $0$ to $2\pi$. The gray-shaded area in
Fig.\,1 corresponds to the region in the complex plane which gets filled by 
the mean eigenvalues of $SU(3)$-matrices. Besides a dihedral symmetry $D_3$
one observes cusps at the cube roots of unity $1,\,(-1\pm i \sqrt{3})/2$ which 
obviously correspond to the center of the compact complex Lie group $SU(3)$: 
$Z(SU(3)) = {\mathbb{Z}}_3$. We can also give a purely algebraic description
of the boundary curve in Fig.\,1. Eliminating $\cos \theta$ from the real part 
of eq.(3) and inserting the solution of that quadratic equation into the
squared imaginary part gives:    
\begin{equation}  Y_b^2 = \pm{2\over \sqrt{3}} (1+2X_b)^{3/2} -1-4X_b-X_b^2 \,,
\end{equation} 
where the signs $\pm$ correspond to the two branches for $-1/2 \leq X_b\leq 
-1/3$ in both the upper and the lower halfplane. After some further elementary 
transformations we arrive at the result that the trace figure of $SU(3)$
(i.e. the locus of its mean eigenvalues) is the region: 
\begin{equation}  (1+3X)(1-X)^3-6Y^2(1+4X+X^2)-3Y^4 \geq 0\,, \end{equation}
inside the unit disc $X^2+Y^2\leq 1$ which is bounded by a $quartic$ curve.  
With the help of the parameterization in eq.(3) we can now compute several
other geometrically interesting properties of the trace figure of $SU(3)$. The 
enclosed area amounts to:  
\begin{equation} \Omega_3  =-2 \int_0^{\pi} d \theta \, {d X_b \over d\theta}
  \,Y_b  = {2\pi \over 9}\,,\end{equation}
and the length of its boundary curve is: 
 \begin{equation} L_3  =6 \int_0^{\pi/3}d \theta \, \sqrt{\bigg({d X_b \over
d\theta} \bigg)^2 + \bigg({d Y_b \over d\theta} \bigg)^2} = {16\over 3} \,.
\end{equation}
The radius of the largest circle that fits into the trace figure, determined by
the minimum of $X_b^2+Y_b^2 = (5+4 \cos 3\theta)/9$, is easily found to be:
\begin{equation} R_3 = {1\over 3} \,. \end{equation}
For the Lie group $SU(3)$ of rank two, one has the special situation that the 
complex number $Z=X+i\, Y= {1\over 3}{\rm tr}\,U$ (subject to the constraint
eq.(5)) represents uniquely a conjugacy class of $SU(3)$. Therefore the 
normalized invariant integral for class functions can be converted into a
two-dimensional integral over the trace figure: 
 \begin{equation}  \int\limits_{SU(3)}\!\! \!\!dU f_{\rm cl}(U)={27\sqrt{3}
\over 2\pi^2  } \int\limits_\Delta\!\!  d^2Z\sqrt{4(1+Z^3+Z^{*\,3})
-3(1+Z^*Z)^2} \, \tilde f(Z,Z^*)  \,. \end{equation}
The square-root type weighting function originates from the Weyl determinant 
\cite{simon} expressed in terms of the coordinates $(X,Y)$ and $\Delta$ is the 
region described in eq.(5), where the radicand is positive. The measure is
$d^2 Z = d X dY$.

\section{Special unitary groups SU(n), $n\geq 4$}
The previous considerations can be straightforwardly generalized to the higher 
special unitary groups $SU(n),\, n\geq 4$. The complex numbers: 
\begin{equation}  X+i\,Y = {1\over n}\, {\rm tr}\, U \,, \quad U \in SU(n)
\,,\end{equation} 
can be rewritten in terms of the $n-1$ (independent) unitary eigenvalues $z_j$ 
as:
\begin{equation}  X+i\,Y = {1\over n}\bigg(z_1+\dots +z_{n-1}+{1\over z_1
\dots  z_{n-1}} \bigg)  \,.\end{equation}
For an extremal boundary point we eliminate all but one angle through the
zero-derivative conditions: $\partial(X+i\, Y)/\partial z_j =0$, $j=1,\dots, 
n-2$ which give $z_1  \dots  z_{n-1}=z_1^{-1}=\dots =z_{n-2}^{-1}$. This 
implies $z_{n-1} = z_1^{1-n}$, and with that we obtain the following
parameterization of the boundary curve in terms of $z_1 = e^{i \theta}$:
\begin{eqnarray}  n(X_b+i\, Y_b) &=& (n-1)z_1 +z_1^{1-n} \nonumber \\
&=& (n-1)\cos \theta +\cos(n-1)\theta +i\, [ (n-1)\sin\theta -\sin(n-1)\theta
] \,.\end{eqnarray} 
The grey-shaded areas in Figs.\,2, 3 and 4 correspond to the regions in the
complex plane filled by the mean eigenvalues of $SU(4)$, $SU(5)$ and 
$SU(6)$-matrices, respectively. In each case one observes in addition to a 
dihedral symmetry $D_n$ cusp points at the $n$-th roots of unity which 
reflect the discrete center subgroup $Z(SU(n)) = {\mathbb{Z}}_n$. The area 
enclosed by the trace figure of $SU(n)$ comes out as a rational multiple of
$\pi$, namely:  
\begin{equation} \Omega_n  = {\pi \over n^2}(n-1)(n-2)\,,\end{equation}
and the circumference of its boundary curve, parameterized in eq.(12), is:
\begin{equation} L_n  = {8\over n}(n-1) \,. \end{equation}
The radius of the largest circle that fits into the trace figure of $SU(n)$, 
determined by minimum of $X_b^2+Y_b^2 = [n^2-2n+2+2(n-1)\cos(n\theta)]/n^2$,
is also readily found to be:  
\begin{equation} R_n = 1-{2\over n} \,. \end{equation}
When extrapolating to large dimensions, $n \to \infty$, one realizes that the 
trace figure tends to fill out the whole unit disc, $\Omega_\infty = \pi$, 
whereas its circumference becomes significantly longer than the unit circle, 
$L_\infty = 8> 2\pi$. This difference arises from the $n$ oscillations of
the boundary curve between $R_{\rm max}=1$ and $R_{\rm min}=1-2/n$.   
In the special case of $SU(4)$ we can furthermore employ the trigonometric 
identities: $3 \cos\theta +\cos 3\theta= 4 \cos^3 \theta$ and $3\sin\theta- 
\sin 3\theta = 4 \sin^3 \theta$, and get the algebraic characterization:  
\begin{equation} |X|^{2/3} + |Y|^{2/3} \leq 1 \,, \end{equation}
of the region filled by the mean eigenvalues of $SU(4)$-matrices. Finally, it 
is interesting to note that the curves parameterized in eq.(12) are the 
so-called hypocycloids. These hypocycloids are constructed by unrolling a 
circle of radius $1/n$ inside the unit circle and recording the motion of a
point on the boundary of the small circle (of radius $1/n$).  
 
\section{Spin groups Spin(4n+2)}
Among the classical simple and simply connected compact Lie groups 
\cite{simon,collins} the spin groups Spin$(4n+2)$ (defined as the two-sheeted
universal covering groups of the special orthogonal groups $SO(4n+2)$) are 
distinguished by the property of possessing a complex fundamental 
representation of dimension $4^n$. The reason for that is that in these 
particular dimensions the (even) real Clifford algebras used to construct the 
spin groups are isomorphic to complex matrix algebras \cite{broecker}. In 
order to analyze the mean eigenvalues: 
\begin{equation}  X+i\,Y = 4^{-n}\,{\rm tr}\, U \,,\quad U \in {\rm Spin}(4n+2)
\subset SU(4^n)  \,,\end{equation} 
for Spin$(4n+2)$ it is sufficient to consider the elements in the maximal 
torus of the form \cite{broecker}:
\begin{equation} U=(\cos \theta_1 + \gamma_1\gamma_2 \sin \theta_1) \dots
 (\cos\theta_{2n+1} + \gamma_{4n+1}\gamma_{4n+2} \sin \theta_{2n+1}) \,.
\end{equation}
The $4n+2$ basis elements $\gamma_j$ obey the anticommutation relations 
$\gamma_j \gamma_k + \gamma_k \gamma_j = -2 \delta_{jk} $ and therefore 
generate the real Clifford algebra $Cl_{4n+2}$. Since only even products occur
for $U\in {\rm Spin}(4n+2)$ in eq.(18) we can rewrite it as: 
\begin{equation} U= (\cos \theta_1 + \tilde \gamma_1 \tilde \gamma_2 \sin
  \theta_1)\dots (\cos   \theta_{2n+1} + \tilde \gamma_{4n+1} \sin 
\theta_{2n+1}) \,, \end{equation}
in terms of the $4n+1$ basis elements $\tilde \gamma_j=\gamma_j \gamma_{4n+2}$ 
which by themselves generate the real Clifford algebra $Cl_{4n+1}$. The 
complexified Clifford algebra in one dimension lower is known to be 
isomorphic to the algebra of complex $4^n\times 4^n$ matrices: ${\mathbb{C}} 
\otimes Cl_{4n} = {\mathbb{C}} (4^n\times 4^n)$. From its $4n$ generators 
$\tilde \gamma_j\in {\mathbb{C}} (4^n\times 4^n)$, $j =1,\dots, 4n$ one can 
construct the matrix $\tilde \gamma_{4n+1} = i \,\tilde \gamma_1 \dots \tilde 
\gamma_{4n} $ which anticommutes with all these $4n$ generators and 
furthermore has the  square $\tilde \gamma_{4n+1}^2=-1$. When taking the trace 
of $U\in {\rm Spin}(4n+2)$ in this matrix representation  of $Cl_{4n+1}$ one 
finds readily that all traces of products of $\tilde \gamma$-matrices vanish, 
except for ${\rm tr}(\tilde \gamma_1\dots  \tilde \gamma_{4n} \tilde 
\gamma_{4n+1}) = {\rm tr}(i)= 4^n i$. This leads to the following
parameterization of the mean eigenvalue of a complex Spin$(4n+2)$-matrix: 
\begin{equation}  X+i\,Y = \prod_{j=1}^{2n+1} \cos \theta_j + i \, 
    \prod_{j=1}^{2n+1} \sin \theta_j \,.\end{equation}
In order to find the boundary of the region inside the unit disc which is 
covered when the $2n+1$ angles $\theta_j$ vary between $0$ and $2\pi$ we
search for the extremum of $Y$ under the condition that $X$ is kept constant. 
The convenient method of the Lagrange multiplier leads to the condition that 
all the $2n+1$ angles $\theta_j$ have to be equal on the boundary of the 
trace figure of Spin$(4n+2)$:   
\begin{equation}  X_b+i\,Y_b = \cos^{2n+1} \theta + i \, \sin^{2n+1}\theta \,.
\end{equation} 
With that knowledge one can also give the algebraic description:  
\begin{equation} |X|^{2/(2n+1)} + |Y|^{2/(2n+1)} \leq 1 \,, \end{equation}
of the region in the complex plane covered by the mean eigenvalues of 
Spin$(4n+2)$. The grey-shaded areas in Figs.\,2, 5, and 6 show these regions 
for the cases Spin$(6)$, Spin$(10)$ and Spin$(14)$, respectively. One observes 
cusps at the fourth roots of unity $\pm1,\, \pm i$ which correspond to the 
discrete center subgroup  $Z({\rm Spin}(4n+2))= {\mathbb{Z}}_4$ \cite{simon}. 
Of course, the wellknown isomorphism Spin$(6)= SU(4)$ also shows up in our
analysis of mean eigenvalues. The area enclosed by the trace figure of
Spin$(4n+2)$ is readily calculated to be: 
\begin{equation} \Omega_n = {\pi\,(2n+1)! \over (4^n \, n!)^2}\,,\end{equation}
(again a rational multiple of $\pi$)  and the circumference of its boundary 
curve (parameterized in eq.(21)) is given by the integral:
 \begin{equation} L_n  =(4n+2) \int_0^{1}dz \sqrt{z^{2n-1}+(1-z)^{2n-1}}
   \,.\end{equation}
Its first four values read:  
\begin{equation} L_1 = 6\,, \qquad L_2= 5 + { 5 \sqrt{3} \over 6} \ln(2+
\sqrt{3}) =6.90086\,,\qquad L_3= 7.43369\,,\qquad L_4= 7.71268\,.\end{equation}
For large $n \to\infty$ the area approaches now quickly zero, $\Omega_\infty = 
0$, whereas the circumference tends to $L_\infty = 8$. The radius of the 
largest circle contained in the trace figure of Spin$(4n+2)$ is: 
\begin{equation} R_n = 2^{-n} \,, \end{equation}
obtained by setting $\theta = \pi/4$ in eq.(21).
\section{Complex exceptional Lie group $E_6$}
Among the five exceptional Lie groups there is exactly one candidate with 
a complex fundamental representation, namely $E_6$ with its (defining) 
27-dimensional complex representation \cite{gourdin}. The mean eigenvalue of 
an $E_6$-matrix is then given by the complex number: 
\begin{equation}  X+i\,Y = {1\over 27}\,{\rm tr}\, U \,,\quad U \in E_6   
\subset SU(27)  \,.\end{equation}
Under the maximal compact subgroup $SU(3) \times SU(3) \times SU(3)$ of $E_6$ 
this 27-dimensional complex representation decomposes as \cite{gourdin}:
\begin{equation} {\bf 27} =  ({\bf 3}\otimes  \overline {\bf 3}\otimes  
{\bf 1})  \oplus ({\bf 1}\otimes  {\bf 3}\otimes  {\bf 3}) \oplus (\overline 
{\bf 3}  \otimes  {\bf 1}\otimes \overline {\bf 3}) \,, \end{equation}
where ${\bf 1}$, ${\bf 3}$ and $\overline {\bf 3}$ denote the singlet, triplet
and anti-triplet representations of $SU(3)$. Since ${\rm tr}\,U$ is nothing but
the character of the defining ${\bf 27}$-representation we can use this
decomposition to express ${\rm tr}\,U$ in terms of products of $SU(3)
$-characters. Introducing the abbreviation $\chi(a,b) = a+b +(a b)^{-1}$ for
the character of the ${\bf 3}$-representation of $SU(3)$ we get for the
eigenvalue sum of an $E_6$-matrix:   
\begin{equation} 27(X+i\, Y)  =\chi(z_1,z_2)\, \chi(z_3^{-1},z_4^{-1})  + 
\chi(z_3,z_4) \,\chi(z_5,z_6) +\chi(z_1^{-1},z_2^{-1})\,\chi(z_5^{-1},
z_6^{-1})  \,. \end{equation}
Here, all six complex variables $z_j =e^{i \theta_j}$ run around the unit
circle. In order to get the boundary of the region covered by $X+i\, Y$, we 
impose the zero-derivative conditions $\partial(X+i\,Y)/\partial z_{1,3,5}=0$.
These allow us in a first step to eliminate half of the variables: $z_2 = 
z_1^{-2}$,  $z_4 = z_3^{-2}$,  $z_6 = z_5^{-2}$. Any further zero-derivative 
condition on the reduced expression for $X+i\,Y$ requires one of the three 
remaining variables to be a cube root of unity. We set $z_3=1$, and get an 
expression which is symmetric under the exchange $z_1 \leftrightarrow z_5$. 
Setting finally $z_1=z_5 = \xi \pm i \, \sqrt{1-\xi^2}$ with $-1\leq \xi \leq 
1$ we find the following parameterization of the boundary 
curve:\footnote{Although some steps in this derivation may seem to be 
ambiguous the main result for the boundary curve eq.(30) has been confirmed by 
detailed numerical investigations.}  
\begin{equation} X_b = {1\over 27} (8 \xi^4 +12 \xi^2+16 \xi-9)\,, \qquad Y_b 
= \pm {8  \over 27} (1-\xi)^2(2+\xi) \sqrt{1-\xi^2} \,, \end{equation}
where the signs $\pm$ correspond to the branches in the upper/lower halfplane.
As an example, the parameter value $\xi=-1$ gives the boundary point $X_b = 
-5/27$, $Y_b = 0$ on the negative real axis. Fig.\,7 shows the trace figure of 
the complex exceptional Lie group $E_6$. Besides the dihedral symmetry $D_3$ 
one observes again cusps at the cube roots of unity $1,\,(-1\pm i\sqrt{3})/2$ 
which reflect the center subgroup of $E_6$: $Z(E_6) = {\mathbb{Z}}_3$ 
\cite{simon}. The area of the grey-shaded region in Fig.\,7 amounts to: 
\begin{equation} \Omega(E_6) = {20 \pi \over 243}\,,\end{equation}
(again a rational multiple of $\pi$) and the circumference of the boundary
curve is given by the (elliptic) integral:
\begin{equation} L(E_6) = {4 \over 9} \int_1^4 ds (s-1) \sqrt{5(s-3)^2+16
    s^{-1}}  = 5.59601 \,.\end{equation}
The radius of the largest circle contained in the trace figure of $E_6$ is:
\begin{equation} R(E_6) = {5 \over 27}\,.\end{equation}
This largest circle meets the three boundary points: $5(1\pm i \sqrt{3})/54$,
($\xi= 1/2$) and $-5/27$, ($\xi=-1$). 

All the other simple and simply connected compact Lie groups $G$ have only
selfconjugate representations \cite{collins}. Therefore  their mean
eigenvalues $X(G)$ are real and confined to the interval $[-1,1]$. In the cases
$G= SU(2)$, $Sp(n)$ (the symplectic groups), Spin$(2n+1)$ and $E_7$ with  
center subgroup $Z(G)= {\mathbb{Z}}_2$ \cite{simon} of order $2$, it follows
immediately from continuity that the whole interval $[-1,1]$ will be covered 
by the mean eigenvalues: $-1 \leq X(G) \leq 1$. The same feature applies also 
to Spin$(4n)$ with center $Z({\rm Spin}(4n)) = {\mathbb{Z}}_2 \times
{\mathbb{Z}}_2$ \cite{simon}.  The lower limit $X(G)=-1$ is in each case 
reached by the negative unit matrix contained in the center of these simple 
and simply connected compact Lie groups. This leaves the exceptional Lie 
groups $G_2$, $F_4$ and $E_8$ with trivial center \cite{simon} as special 
cases since for them the (trivial) lower bound $X(G)=-1$ may not be reached. 
In the next section we determine the proper lower bounds on the mean 
eigenvalues for these three exceptional groups.   
 
\section{Exceptional Lie groups $G_2$, $F_4$ and $E_8$ with trivial 
center}
We start with the exceptional Lie group $G_2$ of rank two. Its lowest
dimensional nontrivial representation is the real ${\bf 7}$-representation 
(i.e. the defining representation of $G_2$) and it decomposes under the 
subgroup $SU(3)$ as \cite{gourdin}:   
\begin{equation} {\bf 7} =  {\bf 1}\oplus  {\bf 3}\oplus  \overline {\bf 3}\,.
\end{equation}
This property allows us to write the mean eigenvalue of a $G_2$-matrix:
\begin{equation} X = {1\over 7} \, {\rm tr}\, U \,, \qquad U\in G_2 \subset
  SO(7)\,, \end{equation} 
in terms of the $SU(3)$-character as follows: 
\begin{equation} 7 X = 1+ \chi(z_1,z_2)+ \chi(z_1^{-1},z_2^{-1}) = 1+ 2 \,{\rm
    Re }\, \chi(z_1,z_2) \,. \end{equation}
From our analysis of $SU(3)$ in section 2 we can take over the inequality: 
$-3/2\leq {\rm Re}\, \chi(z_1,z_2)\leq 3$, (see also Fig\,.1) and deduce from 
it that the allowed range for the mean eigenvalues of $G_2$-matrices is: 
\begin{equation} -{2 \over 7} \leq X(G_2) \leq 1 \,. \end{equation}
We continue with the exceptional Lie group $F_4$ of rank four. Its lowest 
dimensional nontrivial representation is the real ${\bf 26}$-representation
(to be viewed as the defining representation of $F_4$). It  decomposes under 
the maximal compact subgroup $SU(3)\times  SU(3)$ of $F_4$ as:   
\begin{equation} {\bf 26} = ({\bf 8}\otimes  {\bf 1})  \oplus ({\bf 3}\otimes  
{\bf 3}) \oplus (\overline {\bf 3}  \otimes  \overline {\bf 3})\,,
\end{equation}
with ${\bf 8}$ the (real) octet representation of $SU(3)$. This fact allows us 
again to write the mean eigenvalue of a $F_4$-matrix: 
\begin{equation} X = {1\over 26} \, {\rm tr}\, U \,, \qquad U\in F_4 \subset
  SO(26)\,, \end{equation}
in terms of $SU(3)$-characters as: 
\begin{equation} 26X = \chi(z_1,z_2)\,\chi(z_1^{-1},z_2^{-1})-1+ 
\chi(z_1,z_2)\,\chi(z_3,z_4) + \chi(z_1^{-1},z_2^{-1})\,\chi(z_3^{-1},z_4^{-1})
\,  \,, \end{equation}
with $z_j$ four complex variables running on the unit circle. By imposing the 
zero-derivative conditions $\partial X/\partial z_{1,3}=0$ we can eliminate two
variables: $z_2 =z_1^{-2}$, $z_4 = z_3^{-2}$. Then we redefine $z_3 = z_2/z_1$ 
in the reduced expression and the further zero-derivative condition $\partial
X/\partial z_1 =0 $ fixes $z_2=z_1^3$. Finally, we set $z_1= e^{i \theta}$ and
get the quadratic expression:
\begin{equation} 13 X = 2(2+\cos 3\theta )^2-5 \geq -3 \,, \end{equation}
from which one can easily read off the lower bound. Altogether this implies 
that the mean eigenvalues of $F_4$-matrices lie in the interval:   
\begin{equation} -{3 \over 13} \leq X(F_4) \leq 1 \,. \end{equation}
The lower limit value $-3/13$ has also been confirmed in numerical studies
starting from eq.(40).

The most demanding case is that of the exceptional Lie group $E_8$ of rank 
eight. The lowest dimensional nontrivial representation of $E_8$ is the real 
${\bf 248}$-representation, i.e. the adjoint representation on its own Lie 
algebra. It decomposes under the maximal compact subgroup $E_6\times SU(3)$ as 
\cite{gourdin}:    
\begin{equation} {\bf 248} =  ({\bf 78}\otimes  {\bf 1})  \oplus ({\bf 1}
\otimes  {\bf 8 }) \oplus ({\bf 27}  \otimes   {\bf 3})  \oplus (\overline {\bf
  27}  \otimes  \overline {\bf 3}) \,, \end{equation}
where the adjoint ${\bf 78}$-representation of $E_6$ decomposes under  $SU(3)
\times  SU(3) \times SU(3)$ as:
\begin{equation} {\bf 78} =  ({\bf 8}\otimes  {\bf 1}\otimes  {\bf 1})
  \oplus ({\bf 1}\otimes  {\bf 8}\otimes  {\bf 1}) \oplus ({\bf 1}  
\otimes  {\bf 1}\otimes  {\bf 8})  \oplus ({\bf 3}\otimes  {\bf 3}\otimes
\overline {\bf 3}) \oplus (\overline {\bf 3}  \otimes  \overline {\bf 3}
\otimes  {\bf 3}) \,. \end{equation} 
The mean eigenvalue of an $E_8$-matrix: 
\begin{equation} X = {1\over 248} \, {\rm tr}\, U \,, \qquad U\in E_8 \subset
  SO(248)\,, \end{equation}
is proportional to the character of the ${\bf 248}$-representation and the 
decompositions eqs.(28,43,44) give us $248 X$ as a Laurent-polynomial in  
eight (unitary) variables $z_1,\dots, z_8$ with integer coefficients. Setting
all these eight variables equal to each other: $z_j = \xi\pm i\sqrt{1-\xi^2}$, 
$j=1,\dots,8$,  with $-1  \leq \xi \leq 1$, we get the expression: 
\begin{equation} 31 X = 8\xi^6 +24 \xi^5+ 12\xi^4 -8 \xi^3 -6\xi^2+1 \,.
\end{equation} 
The sixth degree polynomial on the right hand side of eq.(46) is shown in
Fig.\,8. One sees that within the interval $-1  \leq \xi \leq 1$ its absolute 
minimum value $-1$ is reached on the left boundary $\xi= -1$. Having found
this lower bound we can conclude that the allowed range for the mean
eigenvalues of $E_8$ is:  
\begin{equation} -{1 \over 31} \leq X(E_8) \leq 1 \,. \end{equation}
We note that one can equivalently eliminate half of the eight variables
through zero-derivative conditions as $z_{j} = z_{j-1}^{-2}$, $j=2,4,6,8$, 
and then go with the remaining four variables onto the diagonal: $z_1=z_3 =z_5=
z_7$. This procedure leads to the same expression as in eq.(46). The
lower limit value $-1/31$ has again been confirmed in numerical studies
starting from the full eight-parameter form of $X(E_8)$. 

This concludes our analysis of the mean eigenvalues for the exceptional Lie
groups $G_2$, $F_4$ and $E_8$ with trivial center. We have derived the
rational, negative lower bounds: $-2/7$, $-3/13$ and $-1/31$, respectively.

\begin{figure}
\begin{center}
\includegraphics[scale=0.62,clip]{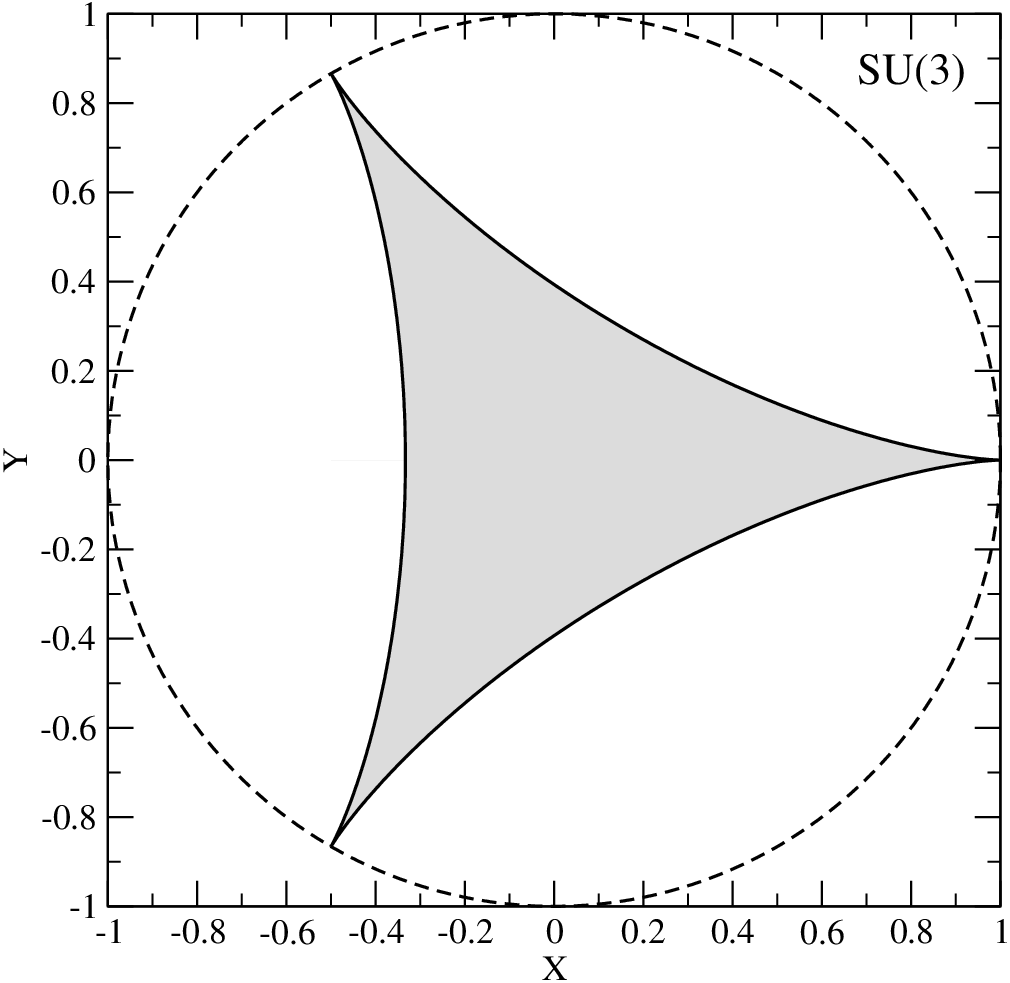}
\end{center}
\vskip -0.5cm
\caption{Region in the complex plane filled by the mean eigenvalues of 
$SU(3)$-matrices. The area inside the (gray shaded) trace figure of SU(3) is 
$2\pi/9$. The boundary curve is a quartic.}
\end{figure}

\begin{figure}
\begin{center}
\includegraphics[scale=0.62,clip]{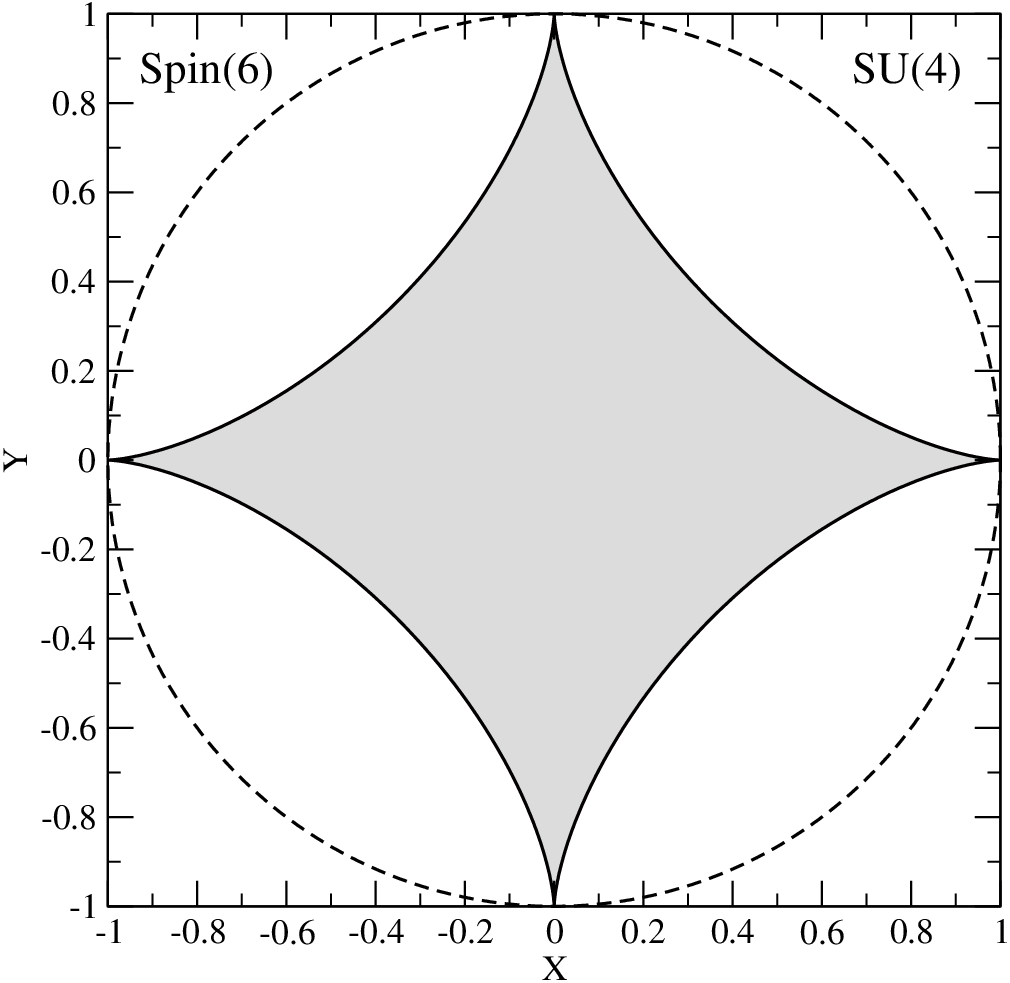}
\end{center}
\vskip -0.5cm
\caption{The trace figure of the complex Lie group $SU(4)={\rm Spin}(6)$. The 
enclosed  area is $3\pi/8$. The boundary curve is a hypocycloid.} 
\end{figure}

\begin{figure}
\begin{center}
\includegraphics[scale=0.62,clip]{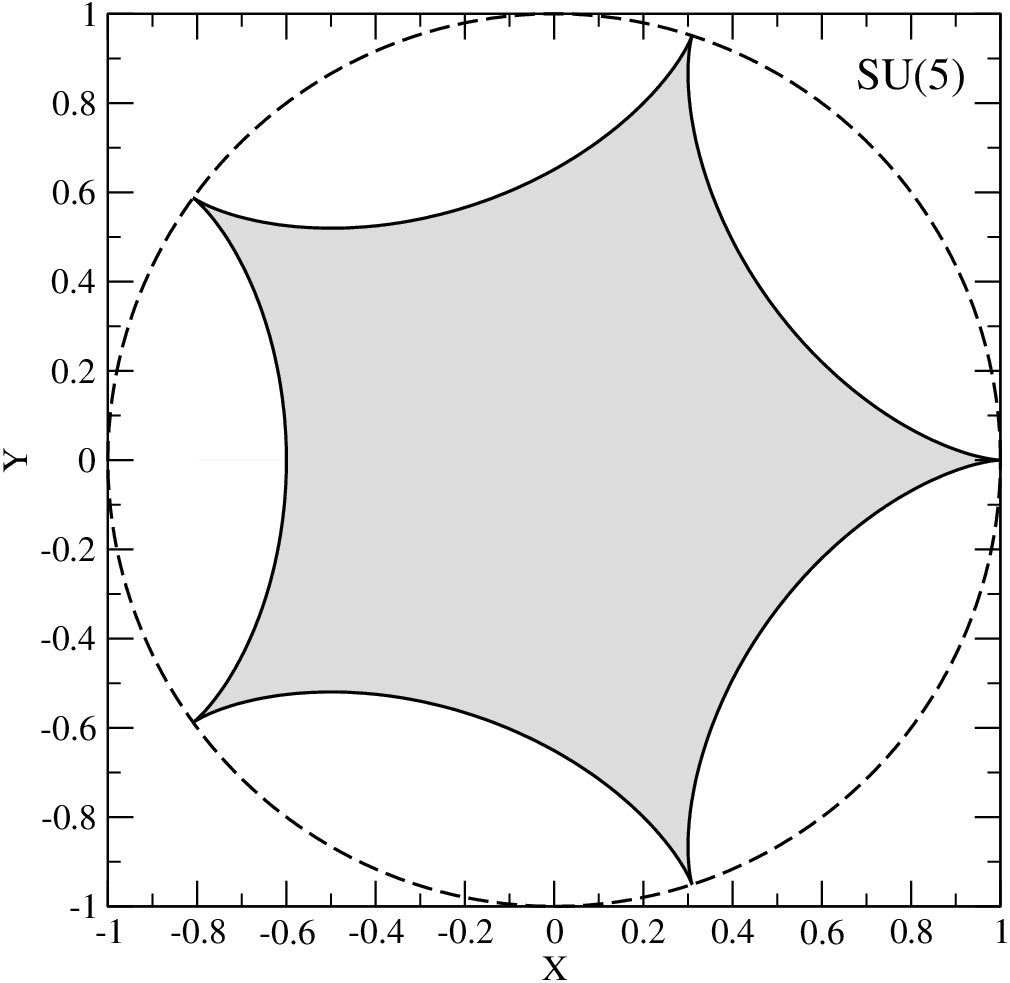}
\end{center}
\vskip -0.5cm
\caption{The trace figure of the complex Lie group $SU(5)$. The enclosed area 
is $12\pi/25$. The boundary curve is a hypocycloid.}
\end{figure}

\begin{figure}
\begin{center}
\includegraphics[scale=0.62,clip]{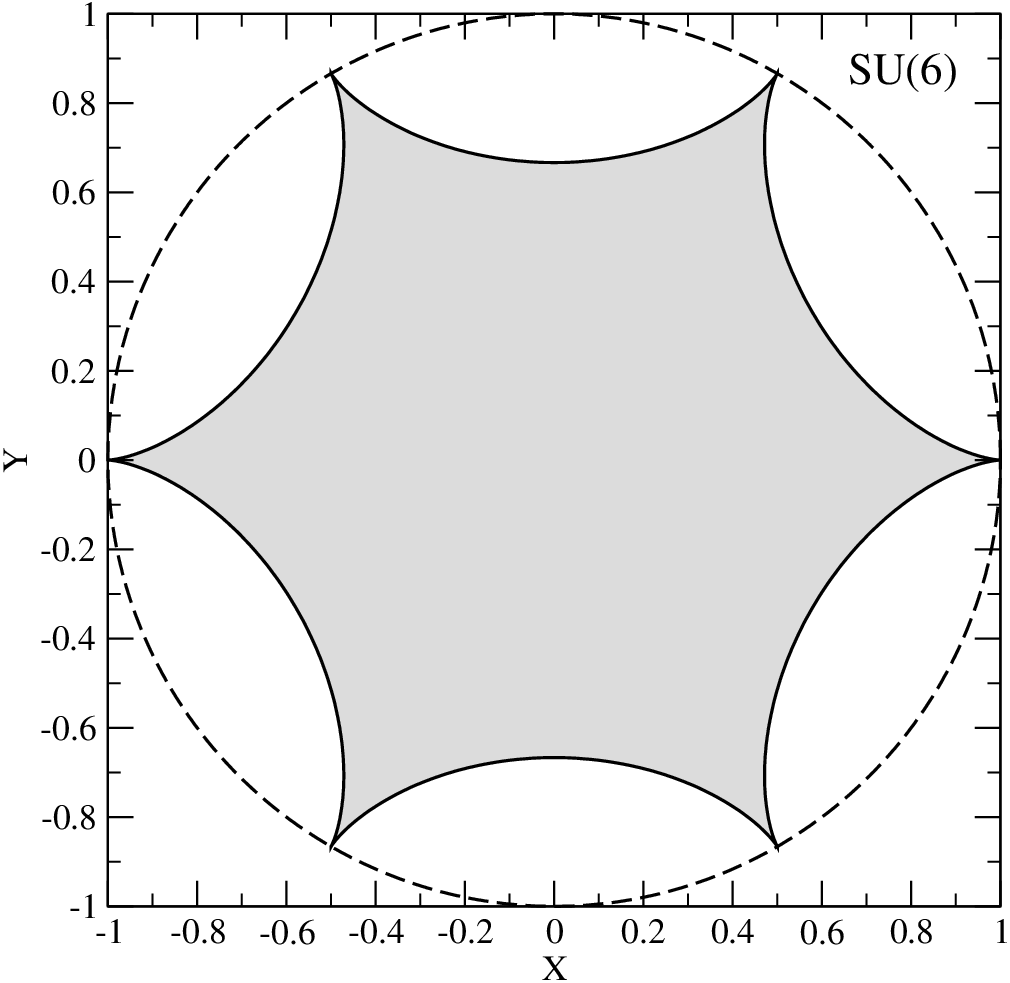}
\end{center}
\vskip -0.5cm
\caption{The trace figure of the complex Lie group $SU(6)$. The enclosed area 
is $5\pi/9$. The boundary curve is a hypocycloid.}
\end{figure}

\begin{figure}
\begin{center}
\includegraphics[scale=0.62,clip]{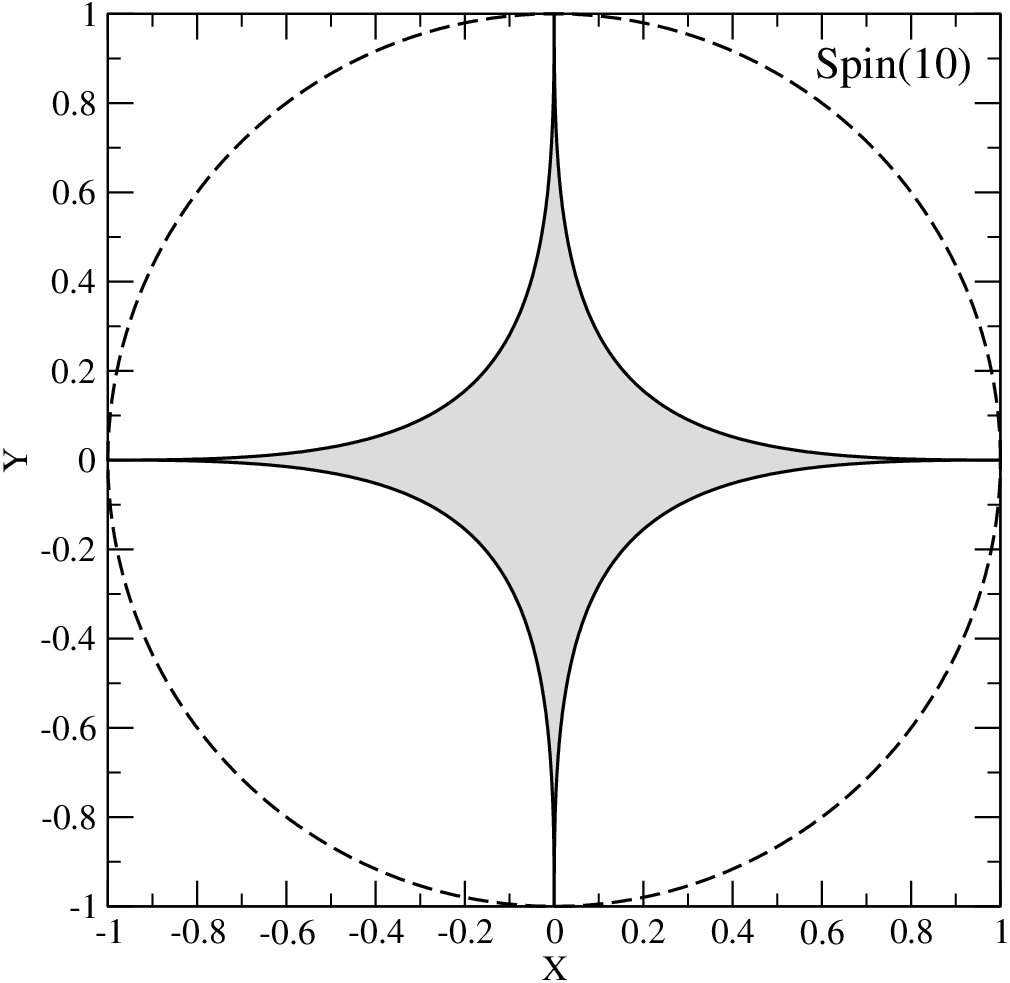}
\end{center}
\vskip -0.5cm
\caption{The trace figure of the complex Lie group Spin$(10)$. The enclosed 
area is $15\pi/128$.}
\end{figure}

\begin{figure}
\begin{center}
\includegraphics[scale=0.62,clip]{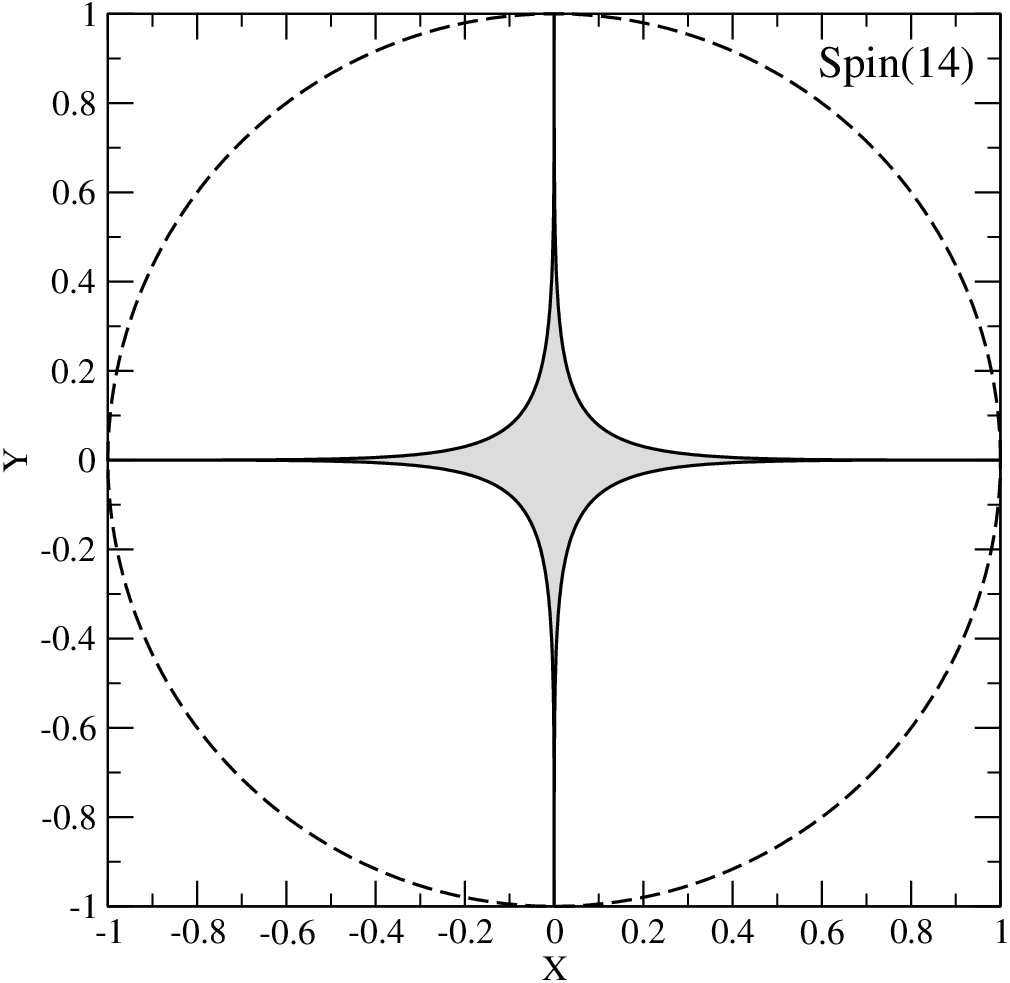}
\end{center}
\caption{The trace figure of the complex Lie group Spin$(14)$. The enclosed 
area is $35\pi/1024$.}
\end{figure}

\begin{figure}
\begin{center}
\includegraphics[scale=0.62,clip]{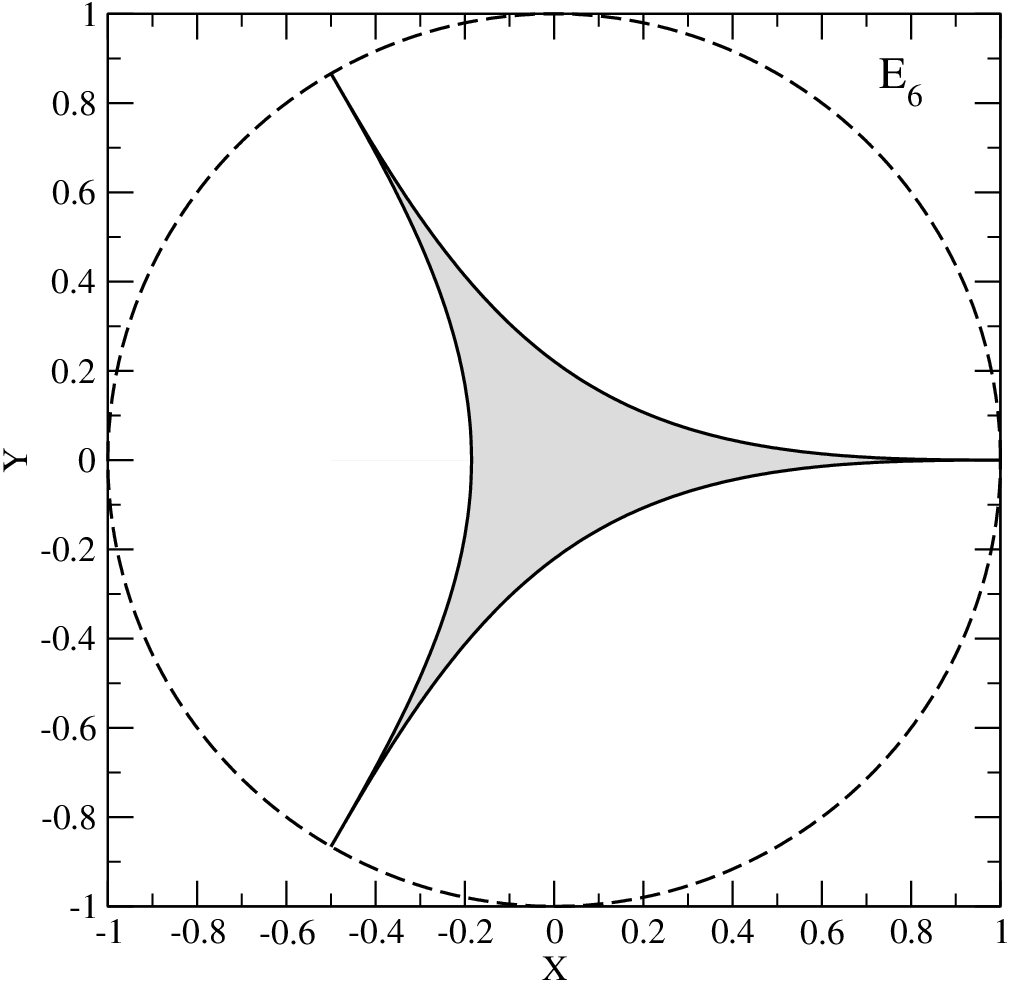}
\end{center}
\vskip -0.5cm
\caption{The trace figure of the complex exceptional Lie group $E_6$. The 
enclosed area is $20\pi/243$.}
\end{figure}

\begin{figure}
\begin{center}
\includegraphics[scale=0.62,clip]{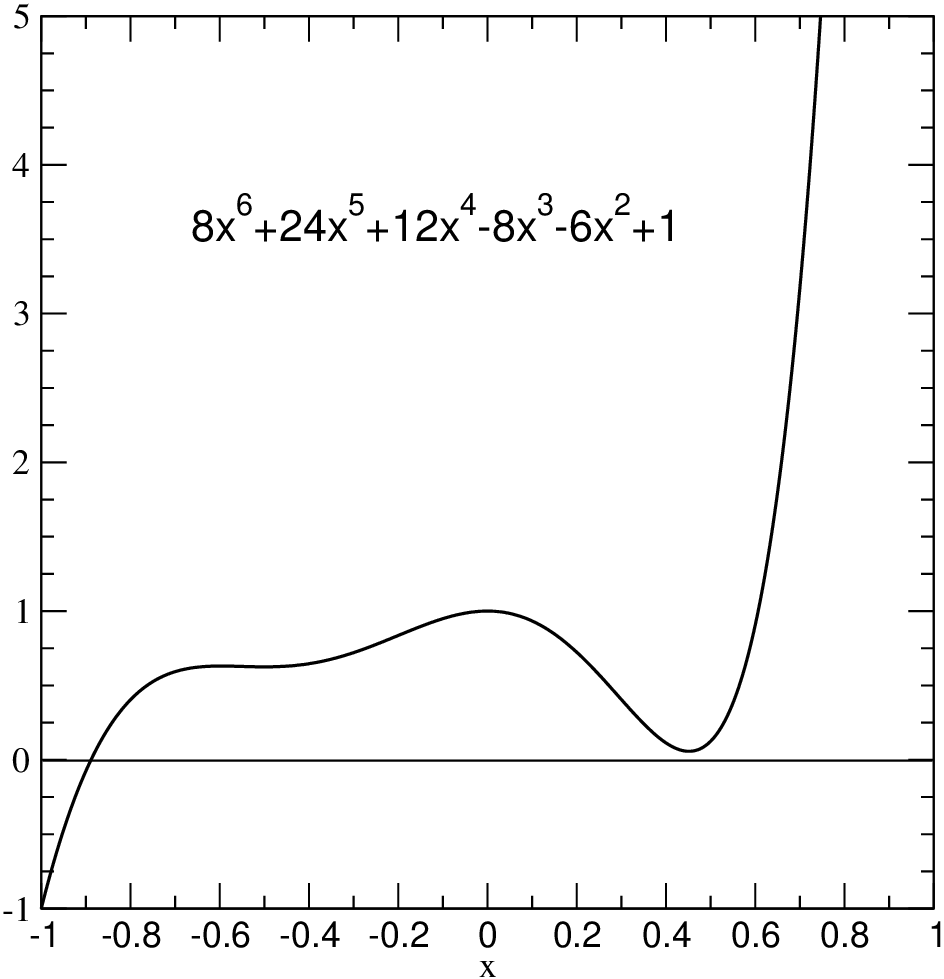}
\end{center}
\vskip -0.5cm
\caption{The polynomial $8\xi^6+24\xi^5+12\xi^4-8\xi^3-6\xi^2+1$ of degree six 
in the  interval $-1 \leq \xi \leq 1$. When multiplied with a factor $1/31$ it
determines the range of the real mean eigenvalues for the exceptional Lie
group $E_8$ in the fundamental {\bf 248}-representation.}  
\end{figure}

\end{document}